\documentclass{iopjournal}

\usepackage{booktabs}
\usepackage{wrapfig}

\begin{document}

\articletype{Paper}

\title{An electron-hadron collider at the high-luminosity LHC}

\author{K D J Andr\'e$^1$\orcid{0009-0005-0028-0852}, L Forthomme$^2$\orcid{0000-0002-3302-336X}, B Holzer$^1$\orcid{0000-0001-8972-2883}, and K Piotrzkowski$^2$\orcid{0000-0002-6226-957X}}

\affil{$^1$CERN, Organisation europ\'eenne pour la recherche nucl\'eaire, Meyrin, Switzerland}

\affil{$^2$AGH University of Krak\'ow, Krak\'ow, Poland}

\email{laurent.forthomme@cern.ch}

\date{\today}% It is always \today, today,

\begin{abstract}
We discuss a concept of a lower-energy version of the Large Hadron-electron Collider (LHeC), delivering electron-hadron collisions concurrently to the hadron-hadron collisions at the high-luminosity LHC at CERN.
Assuming the use of a 20~GeV electron Energy Recovery Linac (ERL), we report the results on the optimised beam dynamics, accelerator technologies, and detector constraints required for such a "phase-one" LHeC.
Finally, we also discuss the ERL configurations and the possibility of delivering electron-hadron collisions during the planned {Run5} of the LHC, which opens excellent research capabilities -- the unique scientific potential of the proposed facility is outlined.
\end{abstract}

%\keywords{accelerator physics, electron-proton collider}
%\submitto{\NJP}
%\maketitle
%\tableofcontents

%%%%%%%%%%%%%%%%%%%%%%%%%%%%%%%%%%%%%%%%%%%%%%%%%%%%%%%%%%%%%%%%%%%%%%%%%%%%%%%%%%%%%%%%%%%%%%%%%%
\section{\label{sec:intro}Introduction}

The Large Hadron-electron Collider ({LHeC}) \cite{LHeCStudyGroup:2012zhm,LHeC:2020van} will open new frontiers in high-energy physics as it will become a unique laboratory for studies of deep-inelastic scattering (DIS) and Higgs boson, electroweak, top quark and beyond the Standard Model ({BSM}) phenomena in a vast range of virtualities $Q^2$, from very close to zero up to $10^6$ GeV$^2$, and for a much extended range of $x$ -- the fractional longitudinal momentum of partons. Recently, an intense electron beam from a novel energy recovery linac (ERL) was proposed to collide with a proton or ion beam from the circular, {27-km} circumference High-Luminosity Large Hadron Collider (HL-LHC) at CERN, in the redesigned P2 interaction region capable of hosting electron-hadron or hadron-hadron collisions, concurrently with hadron-hadron collisions in other {HL-LHC} experiments~\cite{Andre:2022xeh}.

\begin{figure}[h]
    \centering
    \includegraphics[width=0.85\textwidth]{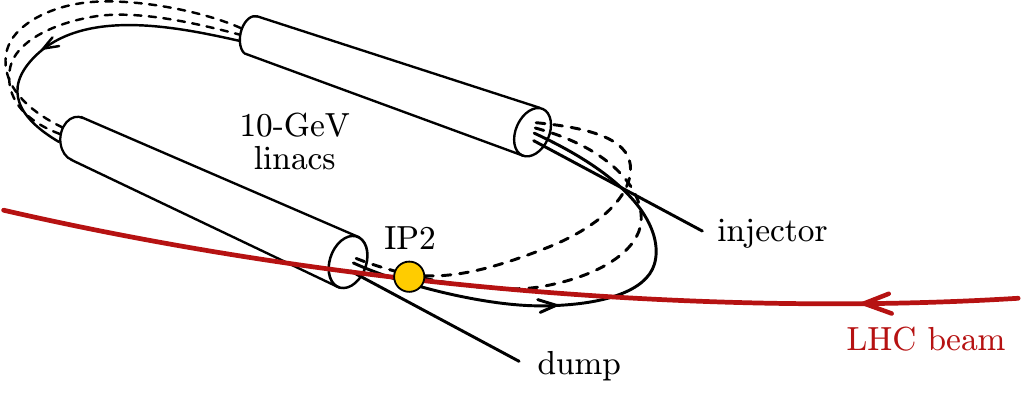}
    \caption{Schematics of the single-pass LHeC layout at the Interaction Point 2 (IP2); with the optional return arcs $3-6$ (dashed lines).}
    \label{fig:layout}
\end{figure}

In the Conceptual Design Report (CDR) of the LHeC, it is assumed that a three-pass ERL delivers a 50 GeV electron beam with an intensity of up to 20~mA~\cite{LHeC:2020van}, and in the most recent update of the LHeC proposal, its commissioning is proposed several years after the completion of the HL-LHC programme \cite{Ahmadova:2025vzd}.
In this paper, we propose the use of a 20~GeV single-pass ERL with a higher beam current of 60~mA, and with a circumference of a third of the LHC's. This choice of design is motivated by the availability of the technology to operate a single-pass ERL (although presently at lower beam currents) and by the overall simplification, in particular, in the design of the adequate machine-detector interface owing to much lower levels of synchrotron radiation, allowing much earlier commissioning of the high energy electron-hadron collisions at CERN. The electron beam will be accelerated to a full energy of 20 GeV by one pass in two 10~GeV linacs (see Fig.\,\ref{fig:layout}), and after collision with the clockwise circulating LHC proton beam, it will be guided back to the linac structures in counter-phase to be completely decelerated within the energy recovery process.\\

\begin{figure}[h]
    \centering
    \includegraphics[width=\textwidth]{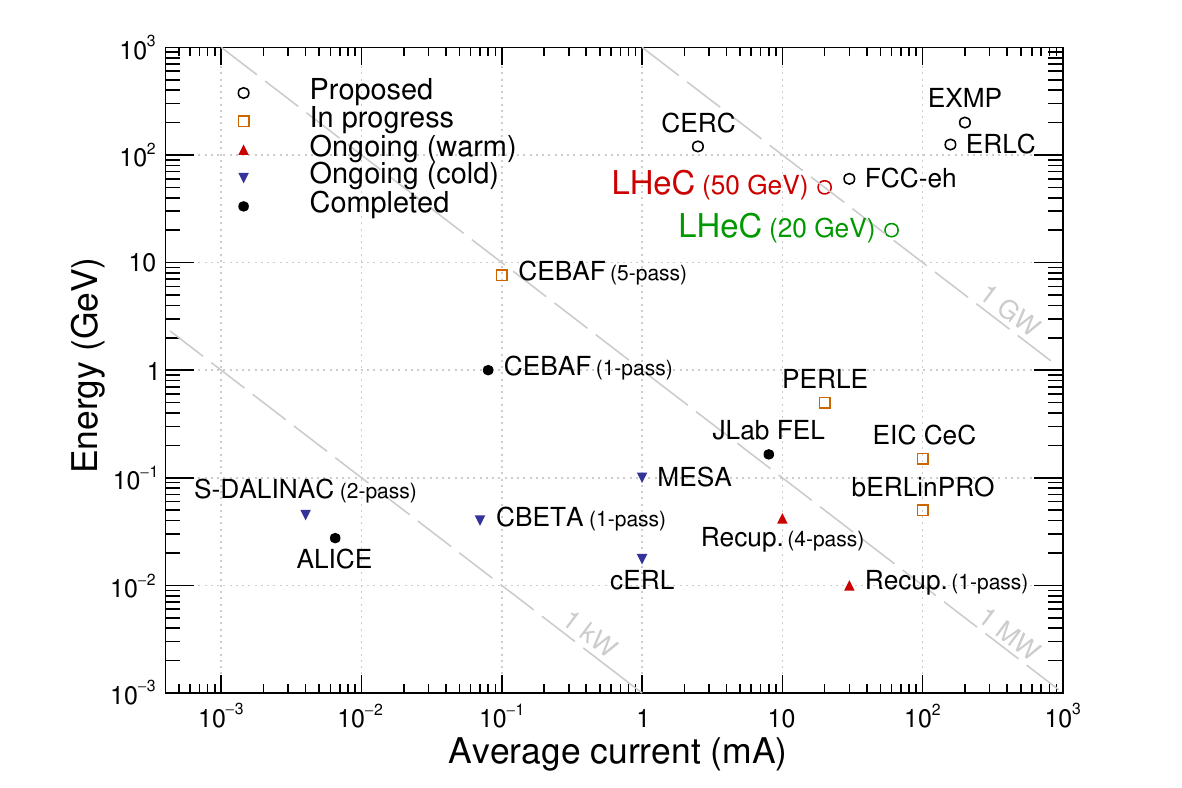}
    \caption{Different ERL projects worldwide that are already operational or in their commissioning / planning phase. The red and green markers indicate the baseline LHeC design and the ERL configuration described in this paper, respectively. Figure adapted from \cite{Hutton:2022kac}.}
    \label{fig:ERLs}
\end{figure}

 The technical feasibility of such a single-pass design has been demonstrated in a number of test facilities and linacs that operate in the range of several mA beam current (see, e.g., \cite{Hutton:2022kac}).
Fig.\,\ref{fig:ERLs} summarises ERLs that are already operational, in the phase of commissioning or planned.

Note that the cost of building this "phase-one" LHeC as well as its running costs are significantly lower than for the final LHeC configuration. Such a staged realisation of the LHeC project provides a huge benefit not only in reducing financial (and personnel) strain but first of all in the availability of concurrent electron-hadron collisions already during {Run5} of the current {HL-LHC} programme. Consequently, apart from the possibility of carrying out a wide range of unique and important studies almost 10 years earlier, the results of this electron-hadron experiment could also be used to improve many important measurements in other (hadron-hadron) experiments at the {HL-LHC} in a timely manner.

%%%%%%%%%%%%%%%%%%%%%%%%%%%%%%%%%%%%%%%%%%%%%%%%%%%%%%%%%%%%%%%%%%%%%%%%%%%%%%%%%%%%%%%%%%%%%%%%%%
\section{\label{sec:interaction-region}The LHeC IR}

The fundamental principle of an ERL provides an ideal condition for a staged particle accelerator.
Still, a number of items have to be followed up, as beam parameters scale with energy, so while keeping the general layout of the ERL and the geometry of linacs and return arcs, a re-optimisation of the beam dynamics is needed for highest performance of the electron-proton collisions.

Schematically, the concept is shown in Fig.\,\ref{fig:layout} -- the geometry of the ERL will remain the same: two superconducting linacs are connected by return arcs on both sides -- details about the general layout can be found in \cite{Holzer:2022eac} and \cite{Bogacz:2022ufs}.
However, for the phase-one case with reduced electron energy several parts needed to reach higher energies are left out: arcs from 3 to 6 will not be installed, nor will the so-called bypass needed in the final 50~GeV design to guide the beam outside the particle detector, which will only be installed in a later staging step.
Beam spreaders and re-combiner modules will also not be present in the phase-one electron beam lattice.

\par On the other hand, a number of key elements will remain unchanged for both ERL phases.
The linacs, the interaction region (IR) and the beam separation scheme will be designed to be valid for any operation energy.
%-----------------------------------------------------------------------------
As described in detail in \cite{Holzer:2022eac,vonWitzleben:2023yca} the proton and electron beams are separated after the interaction point through their different beam rigidities.
A combination of a weak dipole field embedded inside the detector and the combined function effect of the off-centred mini-beta quadrupoles of the electron lattice acts as an effective separation scheme.
The actual value of the dipole strength and the offset of the quadrupoles were optimised for the lowest synchrotron radiation and thus equal effective bending radii in the three magnets \cite{Andre:2022zqx}, resulting in a bending radius of the electron trajectory in this separation scheme of about 1500~m.
%-----------------------------------------------------------------------------

However, while the geometry of the ERL will remain basically untouched, the impact on beam dynamics deserves closer investigation.
Special emphasis is put on:
\begin{itemize}
    \item $p$-optics, orbit;
    \item beam-beam effect on the proton beam / ditto for the electrons;
    \item electron transverse emittance;
    \item beam separation scheme and synchrotron radiation;
    \item design, work load, and cost optimisation;
    \item luminosity reach;
    \item polarisation.\\
\end{itemize}

\begin{figure}[h]
    \centering
    \includegraphics[width=.8\textwidth]{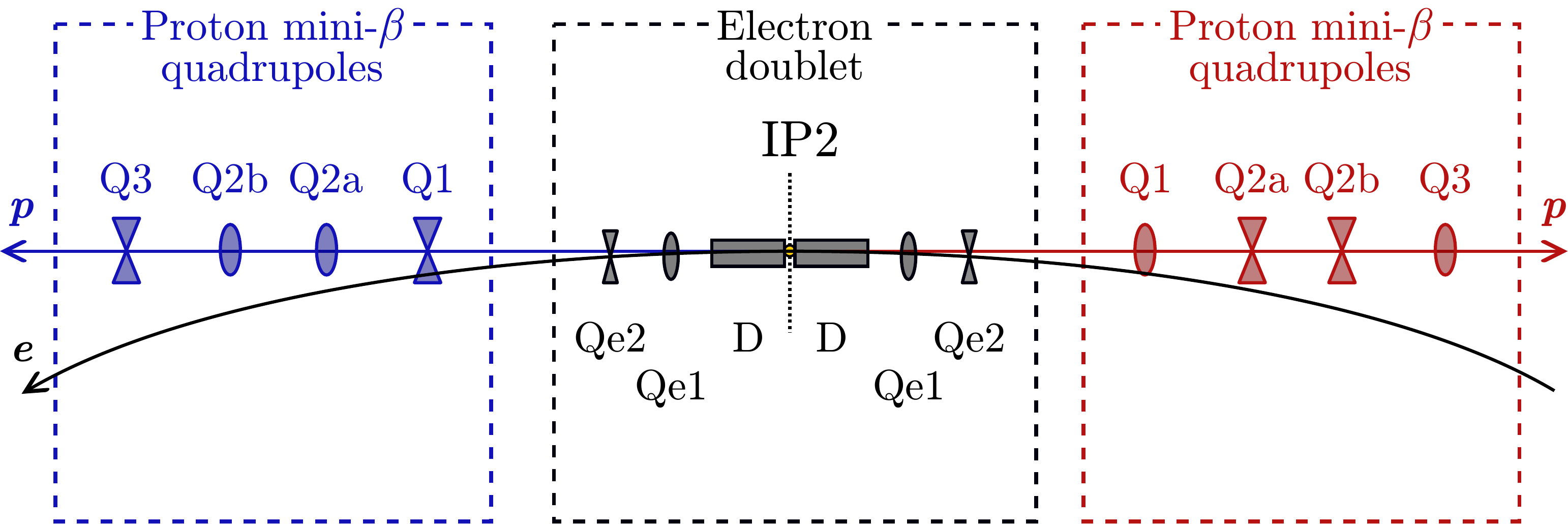}
    \caption{The interaction region of electron and proton beams: in black the mini-beta focusing scheme of the electrons is shown, including the detector dipole field that is part of the beam separation.
    Both are embedded within the free space of the {HL-LHC} proton inner triplet lattice, marked in blue and red. Figure adapted from \cite{tizi}.}
    \label{fig:Tizi}
\end{figure}
The interaction region of the $ep$ collisions is shown schematically in Fig.\,\ref{fig:Tizi}. Given the comparatively low energy of the electrons, a gradient of approximately $g=20~$T/m and a length of 1.8~m are sufficient to meet the optics requirements of the electron quadrupoles. These magnets can therefore be made very compact, and the dimensions of the electron mini-beta insertions have been chosen so that they will fit easily in the free space of the IR. Only the first quadrupole of the proton lattice, ``Q1'' in the figure, will have to be replaced by a new design that allows a field free region for the electrons as their separation is not large enough yet to be guided completely outside the cryostat. For details of the technical design of this new magnet we refer to  \cite{LHeC:2020van}. With the separation scheme explained here, head-on collisions between the proton and electron beam are foreseen, to avoid luminosity reduction because of the geometric loss factor from an eventual crossing angle.
\par
\begin{figure}[h]
    \centering
    \includegraphics[width=.8\textwidth]{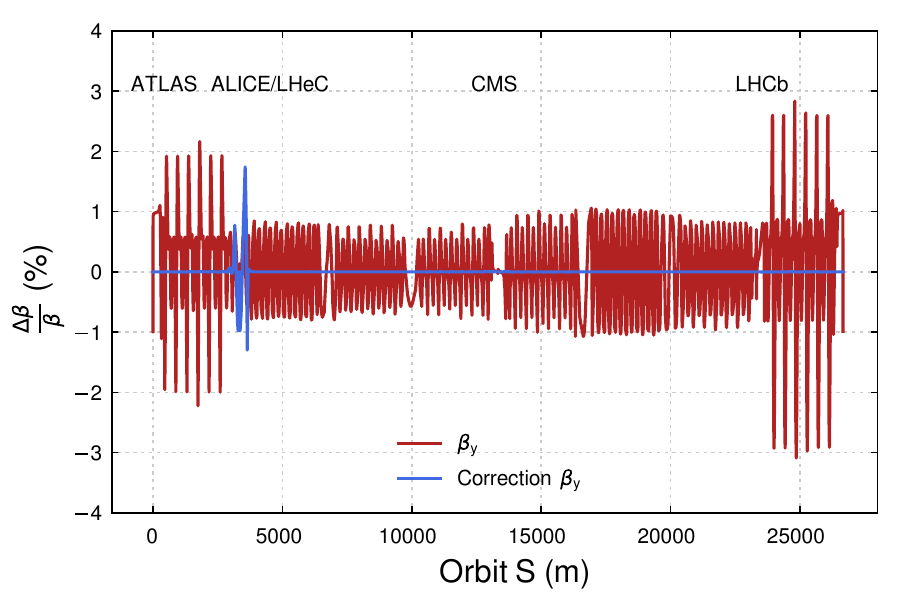}
    \caption{Local distortion of the proton optics due to the influence of the electron quadrupoles before (red) and after a local compensation of the effect (blue) for the case of a 20 GeV electron energy. Figure extracted from \cite{tizi}.}
    \label{fig:p_dist}
\end{figure}

{\bf \boldmath$p$\unboldmath-optics and orbit.} While the electrons will be separated early enough and do not see any influence from the proton mini-beta quadrupoles, the proton beam, on the other hand, will be distorted by the fields of the electron mini-beta quadrupoles.
However, the influence on beam optics and orbit is weak and scales with the ratio of the energy of both counter-rotating beams.
For the design case of the 50~GeV electron and 7~TeV proton beams, the effect has been calculated using a realistic {MAD-X} modelling of the {HL-LHC} lattice including the electron quadrupoles in the IR, and the remaining distortions are compensated.
The so-called beta-beat, $\Delta \beta /\beta$ lies well below the tolerance level for the LHC optics. Nevertheless, it is locally corrected by a rematch of the neighbouring quadrupoles \cite{Holzer:2022eac,vonWitzleben:2023yca}.
Reducing the energy of the electron beam to 20~GeV evidently also reduces this effect, and the calculation of distorted optics, shown in Fig. \ref{fig:p_dist} shows values of the order of $\Delta \beta/\beta = 1.6\% $.
Compared to the overall budget of $\Delta \beta / \beta = 20 \% $ for the tolerance of the LHC optics, this is a negligible contribution; nevertheless, a correction is available and has been applied; see the blue curve in the plot.
\\
\par
{\bf Beam-beam effect on the proton beam.} 
Beyond the optics distortion from the electron gradient fields that have a certain influence on the proton optics, the effect of the beam-beam interaction has been studied for both beams. 
The beam-beam interaction has a highly unbalanced effect, as it depends on the energy and the bunch intensity of the opposing beam.
In linear approximation the resulting tune-shift for the proton beam is given by:
\begin{displaymath}
\Delta Q_{x,y} = \frac{N_e r_0 \beta_{x,y}^\ast}{2\pi\gamma\sigma_{x,y} (\sigma_{x}+\sigma_{y})}
\end{displaymath}
where $N_e$ is the electron bunch population, $r_0$ the classical proton radius, $\gamma$ is the electron Lorentz factor and $\beta^\ast $ and $\sigma_{x,y}$ are the amplitude function of the protons and the transverse beam sizes at the Interaction Point (IP).
We would like to point out that in this approximation the tune-shift does not depend on the $\beta$-function.
However, it depends on the energy of the beam considered.
As expected, due to the limited bunch current of the electrons, the proton beam-beam effect has only a very small impact on the proton emittance and stability. In the end, the electron bunch intensity is in the order of one percent of the proton bunch population, which leads to a very limited beam-beam force compared to the other LHC proton-proton interaction points. The linear part of the beam-beam interaction as well as the non-linear effect has been studied and long-range forces are not present: due to the fast beam separation scheme, the beam separation between electrons and protons is already well beyond 20 $\sigma$ beam size  at the first parasitic bunch crossing and thus -- based on LHC operation experience -- long-range effects can be neglected. Studies on long-term stability, including orbit fluctuations of the electron beam, are ongoing and might support the need for a fast acting correction system.

%For the case of 20 GeV electron and 7 TeV
%proton beams, the effect was estimated using a realistic MAD-X modelling of the 
%{HL-LHC} lattice including the electron quadrupoles in the IR, and the remaining %distortions are compensated; see Fig.\,\ref{fig:p_dist}.
%Due to the low energy of the electrons -- compared to the 7 TeV of the proton %beam -- the insertion quadrupoles of the electron lattice have only a marginal %effect on the proton beam: 
Given the parameters of the LHeC, the resulting beam-beam tune-shift of $\Delta Q_{x,y} \approx 3.8 \cdot 10^{-4}$\ is nearly two orders of magnitude smaller than the effect of the proton-proton interactions for the {HL-LHC} operation where we obtain $\Delta Q_{x,y} \approx -8.6 \cdot 10^{-3}$ per IP and thus the additional distortion due to $ep$ collisions only has a marginal effect.

\par
\begin{figure}[h]
    \centering
    \includegraphics[width=.6\textwidth]{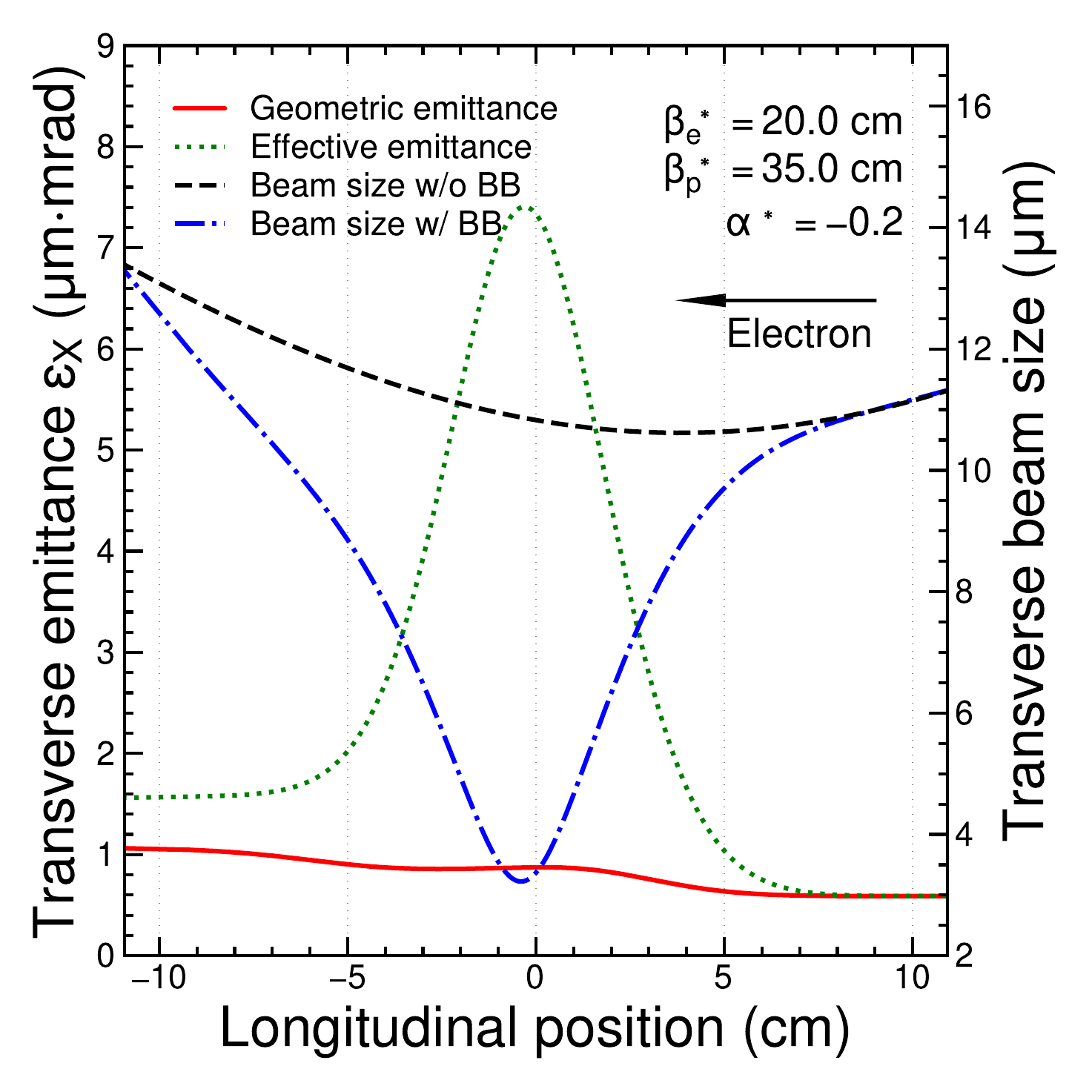}
    \caption{Effect of the space charge force on the 20 GeV electron beam, or ``beam-beam force''. The beam size is reduced (leading to an additional enhancement of the luminosity, pinch effect) and re-matched to the ideal conditions before entering the return arc for energy recovery. The best re-matching was obtained for non-zero value of the slope of amplitude function at the IP, $\alpha^*=-0.2$.}
    \label{fig:el_bb}
\end{figure}

\begin{figure}[h]
    \centering
    \includegraphics[width=.49\textwidth]{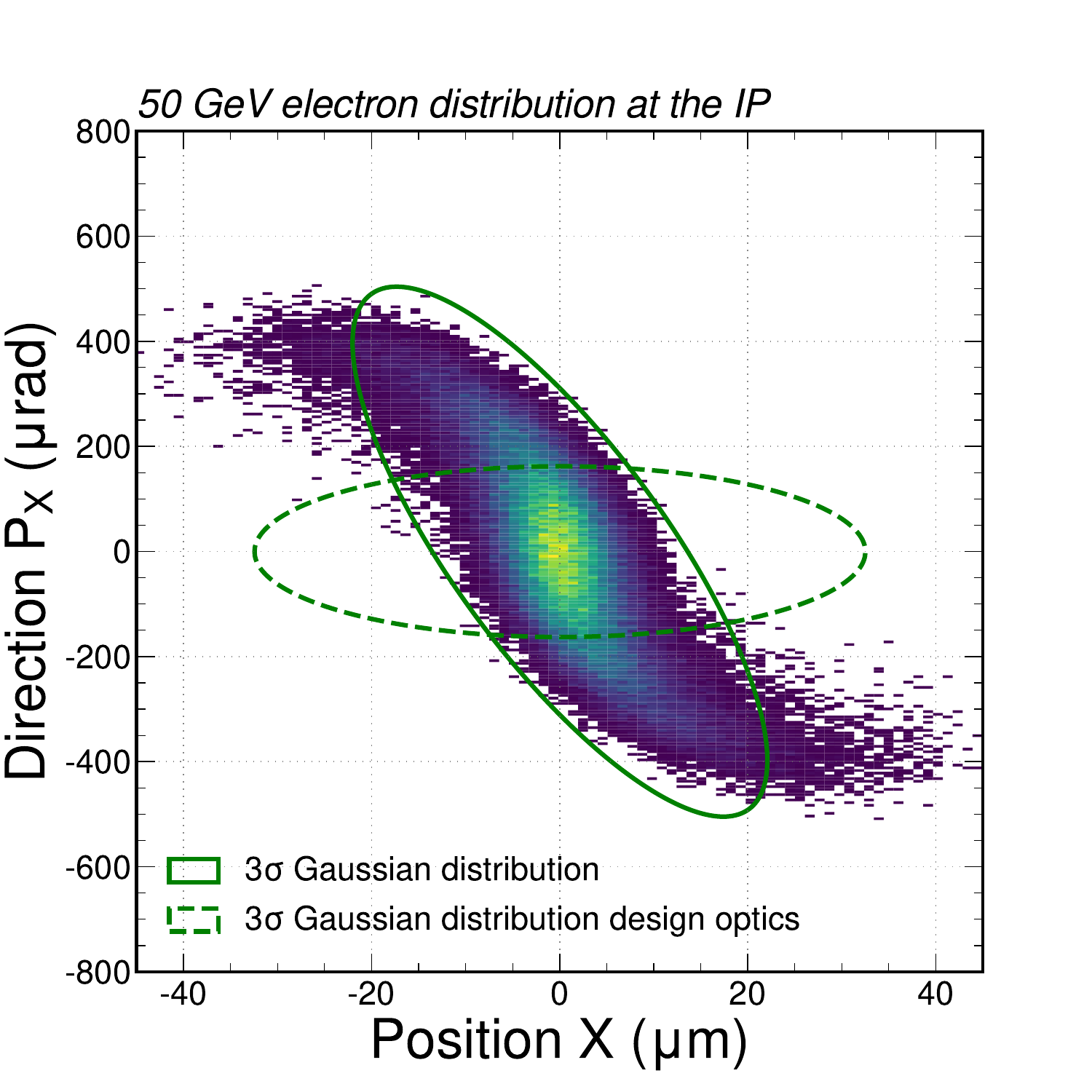}
    \includegraphics[width=.49\textwidth]{fig6a.pdf}
    \caption{Visualisation of the simulated beam-beam effect in phase space: in green the ellipse of the ideal configuration is shown.
    Due to the beam-beam effect, tails in the particle distribution are developing (blue spots) that have to be kept under control for full energy recovery performance.
    Left: 50 GeV case, right: 20 GeV.
    The lower energy of the 20 GeV electrons makes them more susceptible to the beam-beam force and the tails are more developed.}
    \label{fig:e_phase}
\end{figure}

 {\bf Beam-beam effect on the electron beam.} The situation changes considerably for the electrons.
Due to the lower beam energy the electrons are more sensitive, and especially the effect of the non-linear terms has to be minimised.
In Fig.\,\ref{fig:el_bb}, we compute the beam size of the electron beam as a function of the position $s$ along the beam orbit with $s=0$ referring to the IP.
In black, the theoretical beam size is shown as it is obtained without beam-beam forces.
The strong additional focusing due to the space-charge forces of the proton bunch is also shown in blue.
The green dashed line quantifies the mismatch with the design linear optics~\cite{Hao:2014yoa}.
Here we would like to point out that the dominant parameter, the proton bunch population, refers to the ultimate value of $N_p=2.2 \cdot 10^{11}$ as foreseen for the {HL-LHC} luminosity upgrade.
Under the influence of this additional focusing -- called ``pinch effect'' -- the electron beam reaches a smaller effective beam size which is, however, to a large extent compensated by the so-called hourglass effect in the vicinity of the IP.
However, the optics distortion has to be corrected in order to obtain well-matched beam parameters before entering the return arc and the decelerating process of the ERL.
This is also highlighted in Fig.\,\ref{fig:el_bb} where, using the same modelling of the modified {HL-LHC} lattice introduced above, the black and blue curves are shown to be congruent before and after the interaction.
While this fact represents the prerequisite for a successful energy recovery, the non-linear terms of the beam-beam interaction are better visible in the phase-space representation.
We compare the situation for the two electron energies considered here, 50 GeV and 20 GeV, in Fig.\,\ref{fig:e_phase}.
As expected, the beam-beam effects for the lower-energy electrons are more prominent, and therefore stronger tails in the phase space distribution are developing during the beam-beam interaction.
However, still due to the re-matched beam optics the large majority of the particles can be guided back through the decelerating part of the ERL and an energy recovery efficiency of close to 98\% still is obtained.
\par
{\bf Electron transverse emittance.} The above mentioned re-optimisation of the beam-beam interaction involves matched beam sizes between electron and proton beam in both planes:
\begin{displaymath}
\sigma_{x,e} = \sigma_{x,p} = \sigma_{y,e} = \sigma_{y,p}.
\end{displaymath}
Thus, the transverse emittance of the electrons for the 20~GeV case has to be properly considered.
In a storage ring the equilibrium emittance is well defined by the emission of synchrotron light.
In the case of an ERL however, this equilibrium is never reached.
Ignoring the small effect of quantum emission in return arcs, the electron transverse emittance therefore follows the rule of adiabatic damping $\varepsilon \propto 1/\gamma $.
As a consequence, the emittance of the 20~GeV electron beam at the interaction point is considerably larger than the corresponding value at 50~GeV.
This fact has to be taken into account to fulfil the matching condition for the transverse sizes between protons and electrons at the IP.
In principle a strong reduction of the $\beta$-function would be needed to compensate for the increased beam emittance.
However, this is far beyond the capabilities of the mini-beta quadrupoles and the matching of electron beam-size has to be achieved by using the electron source with a reduced transverse emittance.

In addition, the conditions of the matched beam sizes for the 20~GeV phase-one scenario are obtained with the proton beam optics for the LHC Run5, with a $\beta_p^*$ of $35~{\rm cm}$.
Given the emittance and amplitude function of the colliding proton beam, one gets a beam size at the $ep$ interaction point:
\begin{displaymath}
\sigma_p=\sqrt{\varepsilon_p \beta_p^\ast}=\sqrt{\left(3.3 \cdot 10^{-10}~{\rm m}\cdot {\rm rad}\right) \left(35~{\rm cm}\right)}=10.7 ~\mu{\rm m}
\end{displaymath}
which is matched by the electron beam, assuming  an absolute (i.e.~geometric) electron emittance of $\varepsilon_0=5.7\cdot 10^{-4}~{\rm mm\cdot mrad}$ at the energy of 20 GeV which corresponds to a normalised (energy-independent) value of $\varepsilon_n=22~{\rm mm\cdot mrad}$ and lies well within the specifications of the proposed particle source for the ERL, see Tab.~\ref{tab:injector_specification} .  In order to fulfil the requirements of matched beam sizes, a beam optics with $\beta_e^\ast=20$~cm has been established that includes -- as explained above -- the beam-beam effect, namely the additional focusing due to the space charge of the proton bunches, see Fig.\,\ref{fig:el_bb}.
We would like to emphasise in this context that, unlike the proton optics of the {HL-LHC} where a manifold of boundary conditions had to be taken into account, the optics of the final focus system of the ERL is much more flexible.
Consequently, as long as an optics rematch to the return arc 2 is provided to prepare the electron beam for the decelerating phase, a variety of beam optics can be considered.

\begin{table}[!ht]
   \centering
   \caption{General specification of the LHeC ERL electron source~\cite{LHeC:2020van}.}
   %\small
   \begin{tabular}{lcc}
     \toprule
     Parameter & Unit & Value \\
     \midrule
     Booster energy & MeV & 7 \\
     Bunch repetition rate & MHz & 40 \\
     Average beam current& mA & 60 \\
     Bunch charge & nC & 1.5 \\
     RMS bunch length & mm & 3 \\
     Normalised transverse emittance & mm$\cdot$mrad & \textless~6 \\
     Uncorrelated energy spread & keV & 10 \\
     Beam polarisation & \multicolumn{2}{c}{Un-polarised/longitudinally polarised ($\pm 80\%$)}\\
     \bottomrule
   \end{tabular}
   \label{tab:injector_specification}
\end{table}
In these considerations, the small effect of the emittance increase in the return arcs of the ERL was taken into account.
Given the lower beam energy of 20~GeV, the emitted quanta have a much smaller effect, and the resulting emittance dilution remains at a very low level of a few percent. Other effects might influence the quality of the electron beam:
the effect of CSR (coherent synchrotron radiation) has been studied in the context of front-to-end simulations that have been performed for the ERL. Given the parameter choice of the machine no significant influence has been found. 

%Beam-Breakup (BBU) as well as micro bunching instabilities on the other side are items still on the to-do-list towards a technical design report as well as tolerance considerations for the beam-beam effect between the non-colliding proton beam and the electrons, for example. They will be investigated in the near future.

\par {\bf Beam separation scheme and synchrotron radiation.} The beam separation between electrons and protons is provided by the combined effect of the detector weak dipole field and the off-centred quadrupoles of the electron mini-beta concept.
Marked in grey in Fig.\,\ref{fig:Tizi} the combined function quadrupoles of the doublet and the dipole field of the detector represent a bending force with equal and constant bending radius.
They provide a smooth separation effect because of the different beam rigidities of electron and proton beams with a minimised critical energy and limited power of the emitted synchrotron light.
Applying this scheme to the case of 20~GeV electron energy results in $E_{\rm crit} \approx 15-20~{\rm keV}$ and an overall radiated power in the interaction region of well below 1~kW.
Optimisation work is performed modelling the electron lattice including the first {HL-LHC} proton quadrupoles close to the interaction region, and using version 1.7.7 of the beam delivery simulation toolkit (BDSIM) \cite{Nevay:2018zhp},
As for the 50 GeV baseline, two free drift lengths $L^\ast=21$, and $15~{\rm m}$ scenarios are studied in this 20 GeV option for the reduced region of interest centred on the first elements around IP2 (including the first proton quadrupoles), highlighted in grey in Fig.~\ref{fig:Tizi}.
Although the reduction of the radiated power is clear between these two values for the 50 GeV baseline, it is observed to be negligible for the 20 GeV option.
%\par Yet, another slightly modified beam separation scheme is possible: the strong dependence of synchrotron radiation on $\gamma$ allows for the 20~GeV case a much stronger bending as the power of the emitted radiation scales as $P_\gamma\propto{{\gamma^4}/{\rho^2}}$ where $\rho$ is the bending radius of the electrons.
%increased considerably, allowing the electron beam to be guided outside the cryostat of the first proton quadrupole, labelled as {Q1} in Fig.\,\ref{fig:Tizi}.

We would like to point out, however, that in any case a certain contribution of the dipole field to the separation is needed to avoid parasitic encounters of the counter-rotating bunches.
Following the bunch spacing of the LHC, $\Delta t=25~{\rm ns}$, parasitic collisions will occur every 3.75~m and in order to avoid those the beams either have to be brought into collision under a crossing angle (which involves crab cavities to compensate for the geometric luminosity loss) or -- as foreseen in this scenario -- a separation dipole field has to be applied right before and after the collision point, with the total bending power of at least 0.3~Tm.
\par
{\bf Design, work load, and cost optimisation.} Following the idea of a design that can be easily extended in energy by staging additional return arcs, the basic layout of the ERL will remain the same.
Starting with one-pass ERL the return arcs from 3 to 6 as well as the so-called beam spreaders and re-combiners at the end of the linacs and the bypass that guides the low-energy beam outside the particle detector will not be needed.
An estimate of the material cost for the original ERL design resulted in a magnet and vacuum system budget of $11~{\rm kCHF}$ per metre \cite{Bruning:2652349}.
As a result, an overall budget reduction of at least $70~{\rm MCHF}$ is obtained for the 20~GeV version discussed here.
In addition, the work load and associated personnel costs are significantly reduced. Moreover, the lower electron energy allows a much simplified design of the interaction region -- in the baseline case, with $\rho\approx 1500~{\rm m}$, the synchrotron radiation is relatively weak and does not require installation of sophisticated protection masks and absorbers. Finally, as discussed in the following, detector costs are minimised by joining forces with the ALICE 3 project.
%In the alternative design with a faster beam separation, using $\rho\approx 120~{\rm m}$, the corresponding orbit of the electrons can be guided outside the cryostat of the first superconducting proton quadrupole, located at a 23~m distance from the interaction point, and as a result, the layout of the IR in this case is fully compatible with the existing LHC mini-beta scheme, and a new magnet design for the protons (eventually based on ${\rm Nb}_3{\rm Sn}$ technology) is not required for this phase-one scenario.
%The estimated additional cost savings are about $100~{\rm MCHF}$.
Last but not least, the operating costs of the 20~GeV ERL will be much smaller with respect to the 50~GeV case because of much lower energy losses due to synchrotron radiation and the absence of the arc 3--6 beamlines.

As discussed in the 2021 CDR revision of the LHeC study~\cite{LHeC:2020van} the installation of the ERL machine can be made partially in parallel during the Run4 operation period and the detector installation would then take place during the fourth Long Shutdown (LS4) planned in 2035--6 \cite{lhc_run_planning}.
In total, 7 years are needed to build the LHeC, including permits, engineering studies, tendering and land negotiations, and the final year of connection to {HL-LHC}~\cite{Ahmadova:2025vzd}.
Therefore, given a much simpler design of the phase-one LHeC proposed here and a timely decision on the start of such a project, we believe that the electron-hadron collisions can be available during Run5 of the {HL-LHC} which will start, according to the present planning, in 2036.

Last but not least, on the basis of the experience acquired during the construction and commissioning of the phase-one LHeC, the design of the final LHeC will be optimised. Not only at a very technical level, but also in terms of the optimal electron beam parameters and configuration -- for example, by making a well-founded choice between two alternatives: A three-pass ERL setup, providing the highest energy but at the same time a somewhat compromised luminosity, against a two-pass ERL configuration resulting in a lower energy but the highest luminosity.

{\bf Luminosity reach.}
The operational scenario for phase one of the LHeC foresees a truly concurrent running of the $ep$ operation in parallel to the standard {HL-LHC} proton collisions.
As a consequence, the performance of the LHeC depends directly on the {HL-LHC} design parameters, mainly the frequency of bunch collisions $f_{\rm coll}$, bunch populations $N_e$ and $N_p$, and the achievable beta function at the IP.
In the case of round beams and matched beam sizes, to a good approximation $\cal L$ is given by its basic formula:
\begin{displaymath}
{\cal L} \simeq \frac{1}{4 \pi} \frac{N_e N_p }{ \varepsilon_{p}\beta ^\ast_p} f_{\rm coll} .
\end{displaymath}

It should be mentioned that, given the parameters of the LHeC, the usual luminosity reduction factors do not play a significant role -- in particular, head-on collisions between electrons and protons are foreseen at the LHeC. Thus, the luminosity is basically defined by the achievable amplitude function of the proton beam at the interaction point $\beta^\ast_p$ 
which is the leading parameter: in order to achieve highest luminosity, the optics design of the protons in the $ep$ interaction region pushes for minimum possible values of $\beta^\ast_p$, limited by aperture requirements and chromatic budget. The $\beta^\ast$ of the electrons on the other side can be chosen in a wide range, due to the flexibility of the linac lattice that -- unlike the LHC optics -- does not have to observe too many boundary conditions like mini-beta insertions on the other LHC IPs. Accordingly, the combination of $e$-emittance and $\beta^\ast$ of the electron optics can easily follow the matched beam conditions of the beam-beam interaction.

% ---------------------

As a consequence, for a given beam size of the protons at the IP, we assume that the optics of the electrons in the ERL beam delivery system will follow, and the limiting factors are the available free aperture of the proton mini-beta quadrupoles and the dynamic effects created by the sextupole fields needed for chromaticity correction.

The values shown in Tab.~\ref{tab:lumis} summarise the situation, with the nominal proton bunch population, $N_p=2.2\cdot 10^{11} $ and the electron beam current in the ERL, $I_e=60~{\rm mA}$, which corresponds to $N_e\approx 10^{10}$ electrons per bunch. The corresponding values for a 50~GeV electron beam are: $N_e\approx 3\cdot10^{9}$ and $\sigma^\ast= 4.8~\mu$m (assuming, however, the use of the niobium-tin proton mini-beta quadrupoles) \cite{LHeC:2020van}.

% ----------------------------------------------------------
\begin{wraptable}{r}{0.38\textwidth}
\centering
\caption{Luminosity reach in $ep$ collisions computed for an electron beam energy of 20~GeV and the optimal proton optics definition, assuming a 60~mA electron beam current.\\}
\label{tab:lumis}
\begin{tabular}{l cc}
\toprule
 & $p$ & $e^-$ \\
Colliding beam $\beta^\ast$ [cm] & 35 & 20 \\
Emittance $\varepsilon$ [$10^{-10}~{\rm m}\cdot{\rm rad}$] & 3.3 & 5.7 \\
IP beam size $\sigma^\ast$ [$\mu{\rm m}$] & \multicolumn{2}{c}{10.7} \\
Bunch population $N_{p,e}$ [$10^{10}$] & 22 & 1 \\
\midrule
Luminosity [$10^{33}~{\rm cm}^{-2}{\rm s}^{-1}$] & \multicolumn{2}{c}{$6$} \\
Magnet technology & \multicolumn{2}{c}{NbTi} \\
\bottomrule
\end{tabular}
\end{wraptable}
% ----------------------------------------------------------
It should be emphasised that the case described here refers to the truly concurrent operation of electron-hadron and hadron-hadron collisions.
So, the aperture needed in the mini-beta quadrupoles of the protons is defined by both the colliding and the non-colliding proton beam.
Therefore, that non-colliding proton beam still has to be kept in the storage ring and guided through the $ep$ interaction region untouched.
However, in order to save free aperture for the optics of the colliding protons, relaxed beam optics has been developed which allows for minimum beam size and aperture need in the mini-beta quadrupoles: $\beta^\ast =24~{\rm m}$ has been found as optimum value for the non-colliding proton in this case.
\par The same collision scheme will be applied for the other electron-hadron collisions.
For example, assuming the nominal lead beam parameters at the {HL-LHC}, i.e. the beam radius (1 sigma) of 17~$\mu$m and the $Pb$ bunch population of $1.8 \cdot 10^{8}$, one expects the $ePb$ luminosity at the LHeC of $2\cdot 10^{30}~{\rm cm}^{-2}{\rm s}^{-1}$, or about $4\cdot 10^{32}~{\rm cm}^{-2}{\rm s}^{-1}$ per nucleon.

{\bf Polarisation.}
Polarised electron beams can be rather naturally incorporated into an ERL type accelerator. Laser-driven state-of-the-art electron guns provide polarised beams up to 90\% as previously suggested in the LHeC CDR \cite{LHeC:2020van}.
Harmful effects like resonant depolarisation that play a major role in storage rings are by definition not existent, and the de-tuning of the spin precession due to the emitted synchrotron light is estimated to be on the order of a few percent. 
For details see \cite{Bai:1381541}.
As a consequence the acceleration chain provided by the ERL is quasi-spin transparent and does not play a large role in the final -- longitudinal -- electron beam polarisation at the IP.

%%%%%%%%%%%%%%%%%%%%%%%%%%%%%%%%%%%%%%%%%%%%%%%%%%%%%%%%%%%%%%%%%%%%%%%%%%%%%%%%%%%%%%%%%%%%%%%%%%
\section{Physics case}\label{sec:physics-case}

In the phase-one LHeC, the centre-of-mass energy for electron-proton collisions $\sqrt{s_{ep}}=\sqrt{4E_eE_p}=0.75$~{TeV}, in place of 1.2~{TeV} for the 50~GeV electron beam at the final LHeC.
This means that the kinematic reach on the $(x,Q^2)$ plane will already be greatly extended with respect to the data taken at HERA at $\sqrt{s}=0.32$~TeV~\cite{Klein:2008di}, as $Q^2_{\rm max} \simeq s_{ep}$ and $x_{\rm min}\propto1/s_{ep}$.
As discussed in the LHeC CDR~\cite{LHeC:2020van}, a relatively low integrated luminosity of $50~ {\rm fb}^{-1}$ is sufficient to greatly improve the constraints on the parton density functions, particularly at low $x$. The longitudinal electron polarisations of both signs will be instrumental in decomposing the parton distributions, and in particular, in determining the total contribution of sea quarks and their possible charge asymmetry, as well as for measuring the quark density ratio $u/d$. One should stress that it also includes studies of the nuclear parton distributions.
Therefore, already after two years of running the phase one LHeC it will be possible to provide valuable feedback to hadron-hadron experiments at the {HL-LHC}, which will result in significant improvements in a number of precision measurements~\cite{Ahmadova:2025vzd}. 
Moreover, with entering into a new kinematic region of very low $x$ the novel QCD phenomena are expected to be observed -- new parton dynamics and non-linear parton density effects.

\begin{table}[!ht]
\centering
\caption{Inclusive Higgs, single top, and top pair generator level production cross sections for the phase-one and final LHeC, and for un-polarised and polarised electron beams.
In the definition of the process, X corresponds to the dissociative state resulting from the proton interaction.
All cross sections are quoted at leading order, with the Higgs boson mass of $125~{\rm GeV}$ and the top quark mass of $172~{\rm GeV}$.\\}
\begin{tabular}{lrrrr}
\toprule
Proton energy, $E_p$ & \multicolumn{4}{c}{$7~{\rm TeV}$}\\
Electron energy, $E_e$ & \multicolumn{2}{c}{$50~{\rm GeV}$} & \multicolumn{2}{c}{$20~{\rm GeV}$}\\
Electron polarisation, $P_e$ & \multicolumn{1}{c}{$0$} & \multicolumn{1}{c}{$-0.8$} & \multicolumn{1}{c}{$0$} & \multicolumn{1}{c}{$-0.8$}\\
\midrule
$\sigma$($e^-p\to H\nu_l~{\rm +~jet+X}$) [fb] & 73 & 164 & 21 & 52\\
$\sigma$($e^-p\to e^-H~{\rm +~jet+X}$) [fb] & 14 & 19 & 3.4 & 5.6\\
$\sigma$($e^-p\to e^-\,t\bar{t}~+$~X) [fb] & 24.1 & 25.0 & 1.2 & 1.2\\
$\sigma$($e^-p\to \nu_l \,\bar{t}~+$~X) [pb] & 1.5 & 2.6 & 0.28 & 0.50\\
\bottomrule
\end{tabular}
\label{tab:xsec}
\end{table}
\vspace{0.5cm}

To illustrate the scientific potential of the research in electroweak and Higgs physics at the phase-one LHeC, in Tab.~\ref{tab:xsec} we compare four generator-level cross-sections of processes of interest for un-polarised and longitudinally polarised electron beams to those for the final LHeC.
All the values cited here are obtained using the version 3.5.8 of the {MadGraph5\_aMC$@$NLO} \cite{Alwall:2014hca} event generator, using the CTEQ6 \cite{Pumplin:2002vw} set of parton density functions evaluated at leading order.
The Higgs boson case remains compelling at 0.75~TeV (especially for the charged current events), as well as the single-top-quark one, not only for mastering pioneering electron-proton measurement techniques but also for making measurements of some specific channels.
For example, assuming the Run5 total integrated luminosity of $250~{\rm fb}^{-1}$ at 0.75~TeV, and a $-80$\% electron longitudinal polarisation, we obtain the following uncertainties by scaling up the errors obtained for the 1.2~TeV LHeC~\cite{LHeC:2020van} using the numbers of produced events: in the nomenclature of the $\kappa$ framework of an effective field theory (EFT) extension in the Higgs sector \cite{Ghezzi:2015vva}, $\delta\kappa_W\approx2.5\%$ for the Higgs boson coupling to $W$ bosons, and $\delta{\cal R}e(f^2_L)\approx0.2$, for the anomalous $Wtb$ coupling in dimension-6 EFT extensions \cite{Chen:2005vr} -- both errors comparable to the expected final uncertainties for the corresponding measurements at the {HL-LHC} using $3~{\rm ab}^{-1}$ of the $pp$ data.

\begin{figure}
\includegraphics[width=.49\textwidth]{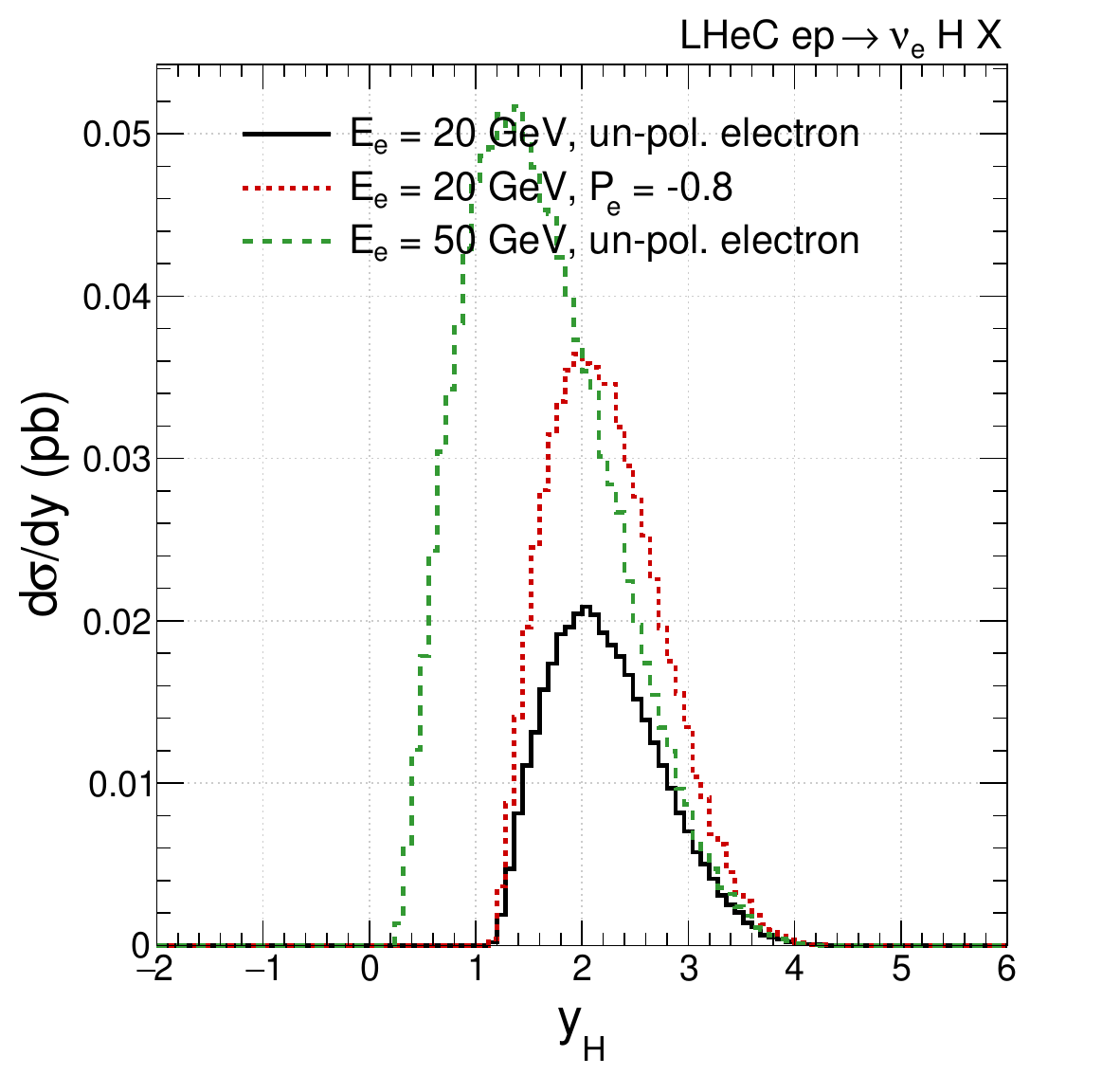}
\includegraphics[width=.49\textwidth]{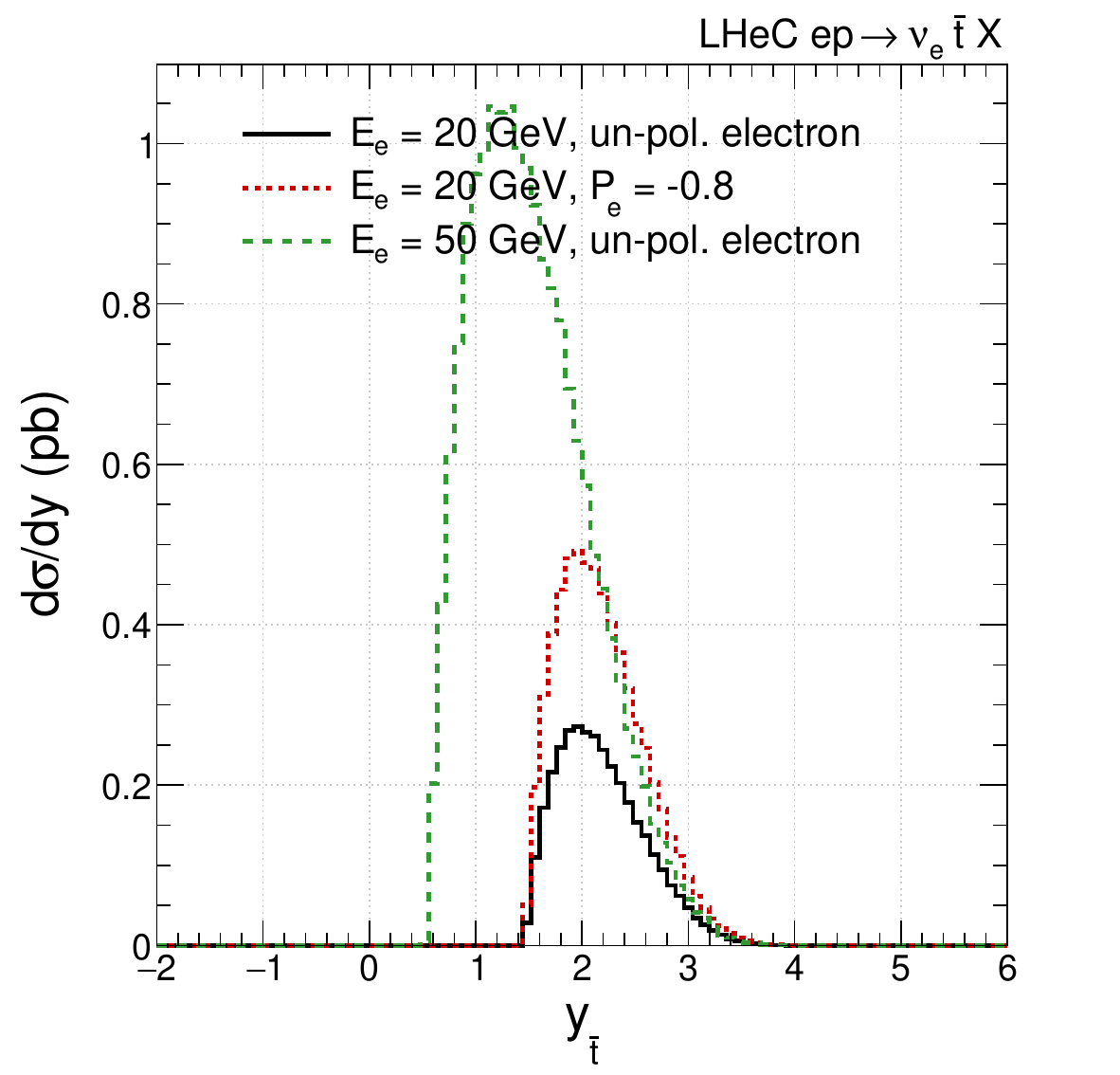}
\caption{Differential cross-sections in rapidity for the Higgs boson (left) and a single top quark (right), both produced through charged current, assuming the phase-one electron beam energy of 20~GeV, with and without electron polarisation, as well as for the final electron beam energy of 50~GeV.}
\label{fig:phys_rapidity}
\end{figure}

In Fig.\,\ref{fig:phys_rapidity} we show the differential cross-sections in rapidity for the Higgs boson and a single top quark produced at the phase-one LHeC as well as the final one -- for the latter the cross-sections peak at about $Y=2$ requiring good coverage in the forward part of the hadron hemisphere.
For the main decay products of these two exemplary processes of interest, a pseudo-rapidity coverage of up to 4 units, both in the tracking and calorimetry components of this hemisphere, would ensure a proper detection efficiency\footnote{The large beam energy asymmetry at the LHeC is well compensated by such a wide $\eta$ coverage, even for the 20 GeV beam -- the electron-parton interactions are kinematically equivalent to the parton-parton interactions
with the ”electron” $x = 20/7000 \simeq 0.003$, which is quite typical at the LHC.}. The big advantage of using the highly polarised electron beam is clearly visible.

Experimental conditions at the LHeC will be particularly favourable for studies of the exclusive production via photon-photon fusion, the event pile-up will be negligible, highly improving the efficiency and purity of exclusive event selection, and at the same time some sources of overwhelming backgrounds, as due to the Drell-Yan process for example, will be absent.
Moreover, no event suppression will occur due to hadronic re-scattering and the associated significant theoretical and experimental uncertainties will not intervene.
Finally, if we consider the usage of data streaming (or the absence of triggering) in acquisition, the detection and reconstruction of a much wider range of final states will be allowed.
As a result, despite the lower average $\gamma\gamma$ centre-of-mass energy than that at the {HL-LHC}, a significantly larger statistics of detected photon-photon interactions is expected already at the phase-one LHeC~\cite{Forthomme:2024qqi}.

For example, using the CepGen v1.2.2 \cite{Forthomme:2018ecc} implementation of the LPAIR \cite{Vermaseren:1982cz} matrix element-level event generator, we estimate a large cross-section for the fully exclusive two-photon production of tau-lepton pairs: about 40~pb at the phase-one LHeC, for pair invariant masses above 10\,GeV.
This will lead to very large event statistics within the acceptance of the central detector and as a consequence to excellent sensitivity to the $\tau$ anomalous magnetic moment $a_\tau$.
With an integrated $ep$ luminosity of 250~fb$^{-1}$, the expected sensitivity is an order of magnitude better than that achieved at the LHC~\cite{CMS:2024qjo}.
As a result, for the first time the experimental uncertainties will be better than the higher-order corrections to the $\tau$ magnetic moment in the Standard Model.
Similarly, large improvements will be achieved in constraining the electric dipole moment of tau-leptons.

Of course, there are many more searches for the BSM physics to be performed at the LHeC as, for example, unique searches for supersymmetric particles and studies of anomalous quartic gauge couplings~\cite{Forthomme:2024qqi}.
Moreover, signatures such as displaced vertices or short tracks may emerge, which can be efficiently identified at the LHeC due to the low level of hadronic background and event pile-up with respect to the LHC, and as such offering a considerable window for discovery. Notable examples involve heavy neutrinos and axion-like particles, inspired by the QCD axion~\cite{LHeC:2020van,Ahmadova:2025vzd}. 

Last but not least, already at the phase-one LHeC, also one of the key questions in nuclear physics can be addressed: how the nucleon structure is modified when immersed in a nuclear medium.
For example, by a precise measurement and complete unfolding of the distributions of all parton species in a single nucleus comparable to that in the proton, the LHeC will revolutionise our understanding of nuclear structure in crucial and unique kinematic regions \cite{LHeC:2020van,Ahmadova:2025vzd}.
In addition, it will allow to clarify the dynamics for particle production in nuclear collisions in the region of relevance for the LHC.
Finally, it should be stressed that the phase-one LHeC provides an important opportunity to perform $eA$ and $AA$ measurements using the same detector which provides the unique advantage
for cross-calibration of performance and physics.

The overlap in the physics programmes of the LHeC and the Electron-Ion Collider (EIC) at Brookhaven is small, as the two projects will produce very important but also very complementary scientific outputs. The maximal centre-of-mass energy of electron-proton collisions at the EIC is 0.14~TeV therefore the electroweak and Higgs physics play there only marginal role~\cite{AbdulKhalek:2021gbh}, in stark contrast to the LHeC. Moreover, the significantly higher centre-of-mass energy at the LHeC corresponds to a much wider coverage on the $(x, Q^2)$ plane, as explained above, allowing for unique studies of non-linear effects in partonic interactions and for significantly improved predictions of low-$x$ processes in the $pp$ collisions at the LHC. On the other hand, both beams can be polarised at the EIC and that will provide means for new insights into the origin of spin in hadrons, in particular. Last but not least, there is a great synergy between the two projects in the detector developments and of the associated electronics.
%%%%%%%%%%%%%%%%%%%%%%%%%%%%%%%%%%%%%%%%%%%%%%%%%%%%%%%%%%%%%%%%%%%%%%%%%%%%%%%%%%%%%%%%%%%%%%%%%%
\section{General detector requirements}\label{sec:detector}

In 2022, an extension of the {ALICE} experiment beyond Run4 at the {HL-LHC} was proposed in a letter of intent \cite{ALICE:2022wwr}.
In that document, two configurations of magnetic field were considered for the new ALICE 3 detector at the IP2 -- produced either by the cost effective 7.5~m long solenoid, or by the high-performance combination of a 2.5~m solenoid at the centre with two 2~m long dipoles on its each side.
As discussed above, to allow for the optimal electron-hadron collisions, the dipole field is necessary for the separation of the electron beam, so the latter approach is desirable in this context.

The most recent {ALICE 3} design includes a tracking detector with excellent performance that covers a wide range of rapidities of $|\eta|<4$ \cite{Dainese:2925455}, which is well compatible with the needs for electron-hadron measurements at the LHeC with 20 GeV electrons.
However, in contrast, the present design lacks a hadronic calorimeter (HCAL) while a performant one is obligatory for performing proper measurements of the DIS charged current events and reconstruction of the decays of top quark, Higgs, $W$ and $Z$ bosons.
Moreover, an extension of the electromagnetic calorimeter is desirable in the electron forward direction to ensure a good reconstruction of DIS neutral current events; in addition, important studies of exclusive production at the LHeC would largely benefit from specialised near-beam detectors in very forward electron and proton directions.

Finally, it should be stressed that the {ALICE 3} design already contains a full data-streaming mechanism (i.e. a "trigger-less" data acquisition), which is also very desirable for measurements at the LHeC.

In summary, it seems that the proposed ALICE 3 detector requires relatively small modifications, apart from the addition of HCAL, to make it also appropriate for the phase-one LHeC experiment at the IP2 -- in this way, allowing vastly increasing the scientific scope and impact of research performed using the Run5 data.

%%%%%%%%%%%%%%%%%%%%%%%%%%%%%%%%%%%%%%%%%%%%%%%%%%%%%%%%%%%%%%%%%%%%%%%%%%%%%%%%%%%%%%%%%%%%%%%%%%
\section{Conclusions}\label{sec:conclusions}

In this paper, we argue that a phased realisation of LHeC offers the best approach.
Firstly, in this case the project financial as well as work load will be distributed over more years.
Secondly, the experience acquired during the first stage of the project will help to make an optimal design of the final LHeC and possibly also allow its earlier high-performance start-up. 
% ------
For example, there are two technical items that will be studied and optimised during the phase one:
\begin{itemize}
\item The performance of an ERL strongly depends on the understanding and control of the RF-phase during the acceleration and deceleration stages. Unlike for a synchrotron, where the synchronous phase is automatically adjusted, phase stability is of crucial importance here.
\item Experience at the SLAC and HERA colliders has shown that orbit fluctuations at the IP have to be damped to avoid emittance blow-up of the beams. The LHC itself is equipped with an orbit stabilisation system to avoid these effects, and experience will show whether additional systems will be needed for $ep$ collisions, e.g. multilayer orbit control as used in high-performance light sources or free-electron lasers.
\end{itemize}
% ------
The exploratory studies presented above demonstrated the high luminosity performance of such a phase-one LHeC. This in turn guarantees an excellent scientific output already during Run5: firstly, it will ensure important and timely scientific feedback to hadron-hadron experiments in HL-LHC as well as unique studies of new phenomena due to high density of partons; secondly, it will allow for making interesting contributions to the studies of electroweak and top physics, for example, and for performing original searches for the BSM physics; and finally, it will allow unique comparative studies of the $eA$ and $AA$ interactions, as both types of collisions are allowed by the proposed IR design.

The new detector at the P2 site can be designed by extension of the {ALICE 3} proposal, which offers strong synergies and benefits in several areas.

%%%%%%%%%%%%%%%%%%%%%%%%%%%%%%%%%%%%%%%%%%%%%%%%%%%%%%%%%%%%%%%%%%%%%%%%%%%%%%%%%%%%%%%%%%%%%%%%%%

\data{No new data were created or analysed in this study.}

\ack{We thank T. M. P. von Witzleben for her contributions to the study of the LHeC optics.}

\funding{L. Forthomme and K. Piotrzkowski appreciate the financial support of the Polish National Agency for Academic Exchange (NAWA) under grant number BPN/PPO/2021/1/00011.
L. Forthomme has also been supported by the NCN grants 2022/01/1/ST2/00022 and 2023/49/B/ST2/03273, and by the AGH IDUB and the CERN COAS programmes.}

\bibliographystyle{iopart-num}
\bibliography{bibliography}% Produces the bibliography via BibTeX.

\end{document}